\documentclass[twocolumn]{aastex63}
\usepackage{amsmath}
\usepackage[T1]{fontenc}
\usepackage{upgreek}
\usepackage{graphicx}
\usepackage{soul}
\received{}
\revised{}
\accepted{}
\submitjournal{ApJ}

\newcommand{\fk}[1]{}
\newcommand{\alfven}{Alfv\'{e}n}

\shorttitle{Strong outflows in ULXs}
\shortauthors{Kayanikhoo et al.}

\begin{document}
% \title{Collimated outflows in moderately magnetized neutron star ULXs}
\title{ULX collimation by outflows in moderately magnetized neutron stars}
%\title{Moderately magnetized accreting neutron stars as ULXs with strong outflows}
%Ultraluminous X-ray sources}
% \title{The impact of magnetic field on the apparent luminosity of neutron star ultraluminous X-ray sources}
%\title{The impact of magnetic field on the apparent luminosity of neutron star ULXs in Radiative GRMHD simulations}
% \title{Impact of the strength of magnetic field on beamed emission in accreting neutron stars as ULXs}

% \title{Beamed emission of accreting neutron stars: Investigating the impact of magnetic field}
\correspondingauthor{Fatemeh Kayanikhoo} 
\email{fatima@camk.edu.pl}

\author[0000-0003-2835-3652]{Fatemeh Kayanikhoo}
\affiliation{Nicolaus Copernicus Astronomical Center of the Polish Academy of Sciences, Bartycka 18, 00-716 Warsaw, Poland}
\affiliation{Research Centre for Computational Physics and Data Processing, Institute of Physics, Silesian University
in Opava, Bezru\v{c}ovo n\'am.~13, CZ-746\,01 Opava,
Czech Republic} 

%%\affiliation{Nicolaus Copernicus Astronomical Center of the Polish Academy of Sciences, Bartycka 18, 00-716 Warsaw, Poland}

\author[0000-0001-9043-8062]{W\l odek Klu\' zniak}
\affiliation{Nicolaus Copernicus Astronomical Center of the Polish Academy of Sciences, Bartycka 18, 00-716 Warsaw, Poland}

\author[0000-0002-3434-3621]{Miljenko \v{C}emelji\'{c}}
\affiliation{Nicolaus Copernicus Superior School, College of Astronomy and Natural Sciences, Gregorkiewicza 3, 87-100, Toru\'{n}, Poland}
\affiliation{Research Centre for Computational Physics and Data Processing, Institute of Physics, Silesian University
in Opava, Bezru\v{c}ovo n\'am.~13, CZ-746\,01 Opava,
Czech Republic} 
\affiliation{Nicolaus Copernicus Astronomical Center of the Polish Academy of Sciences, Bartycka 18, 00-716 Warsaw, Poland}
\affiliation{Academia Sinica, Institute of Astronomy and Astrophysics, P.O. Box 23-141,
Taipei 106, Taiwan}

\begin{abstract}
We perform radiative magnetohydrodynamics simulations in general relativity (GRRMHD) of super-Eddington disk accretion onto neutron stars endowed with a magnetic dipole corresponding to surface strengths not exceeding 100 GigaGauss. 
Accretion is found to power strong outflows which collimate the emergent radiation of the accretion columns, leading to apparent radiative luminosities of $\sim 100$ Eddington, when the true luminosity is a few Eddington units. Surprisingly, the collimation cone/angle widens with increasing magnetic field. Thus, in our simulations the apparent luminosity of the neutron star is substantially larger for the weaker magnetic fields ($10^{10}\,$G) than for the stronger ones ($10^{11}\,$G). We conclude that a super-Eddington accreting neutron star with the dipole magnetic field $10^{10}\,$G is the most likely source of ultraluminous X-rays. 

\end{abstract}


\section{Introduction}

Ultraluminous X-ray sources (ULXs) have puzzled astrophysicists since the 1990s. These non-nuclear extragalactic sources emit X-rays at luminosities exceeding $10^{41}\, {\rm erg \,s^{-1}}$, which is less than the luminosity of Active Galactic Nuclei (AGNs) but far surpassing the Eddington luminosity ($L_\mathrm{Edd}$) for a typical stellar-mass black hole (of mass $\sim10\, M_\odot$) or a neutron star. Such extraordinary luminosity may be produced in binary systems where the accretor is a stellar-mass black hole or a neutron star that apparently emits beyond its Eddington limit \citep{King2001}. 

Over the years, different models have been proposed to explain ULXs. \cite{Colbert1999} suggested that the high luminosity of ULXs was attributed to sub-Eddington accretion in intermediate-mass black holes (IMBHs) with masses in the range of $10^2$ to $10^4\,M_\odot$. \cite{Begelman_2002} suggested photon-bubble instability in accretion onto stellar mass black holes, but the model cannot explain luminosities higher than $3\times 10^{40}\,{\rm erg\,s^{-1}}$ without beaming \citep[see][and references therein]{lasota2024problems}.

\cite{King2001} noted that ULXs may represent a transient stage in high-mass X-ray binaries (HMXB) characterized by extremely high mass transfer rates with compact accretor: stellar-mass black hole ($\sim 10M_\odot$), neutron star or white dwarf. They proposed super-Eddington accretion onto compact objects in intermediate and high mass X-ray binaries and emission geometrically beamed by outflows close to the accretor that create a funnel-like optically thin region. Radiation can reach the observer through the funnel, which is a fraction $b\ll1$ of the solid angle of the sphere. Such an observer overestimates the true luminosity $L$ by a beaming factor $b$ related to the apparent (so-called isotropic) luminosity; $L_{\rm iso} \sim L /b$. Our simulations support this scenario.  

We note that super-Eddington accretion and radiatively driven disk outflows were considered for black holes already by \cite{Sunyaev1973}, who find the 
disk luminosity to be proportional to $\sim 1 + \ln (\dot{M}/\dot{M}_{\rm Edd})$. Here $\dot{M}/\dot{M}_{\rm Edd}$ is the mass accretion rate in the units of $L_{\rm Edd}/(\eta c^2)$, the Eddington luminosity divided by a suitable efficiency factor. The emitted disk radiation exceeds the Eddington limit within a specific radius called the spherization radius. However, in the case of neutron stars the stellar magnetic field may significantly affect the outflows, and the luminosity of the accretion stream hitting the stellar surface is expected to be on the order of $L\sim GM\dot M/R_*$, i.e., $L\sim 0.2\dot M c^2$ for a stellar radius of $R_*=5 r_{\rm g}$, where $r_{\rm g}=GM/c^2$ is the gravitational radius.
In accreting neutron stars, the magnetic field truncates the accretion disk, allowing the material to accrete onto the neutron star along the magnetic field lines anchored in the star. Radiation can escape from the sides of the column, as detailed by \cite{basko1976}. If the spherization radius is larger than the magnetosphere the disk may have radiatively driven outflows between the two radii.                 

In 2014, \cite{bachetti2014} discovered that the source ULX–2 in the galaxy M82 exhibits pulsations in its lightcurve with an average period of 1.37~s. The coherence and the short period of this source indicate that the central object is a neutron star. This neutron star must accrete beyond Eddington limit, its apparent luminosity is $\sim 10^3\, L_\mathrm{Edd}$ in the pulsed emission alone. This finding increased the possibility that ULXs are powered by super-Eddington accreting neutron stars. Subsequently, several other pulsating ULXs were discovered with spin periods and their derivatives in the range of less than a second to 12 minutes \citep[e.g., ][]{Motch_2014,Israel2016, Furst2018, Chandra2020}.

In some models, magnetars are considered as the accretor in ULXs \citep{Eksie2018,Mushtukov2018}. Magnetic fields $\gtrsim 10^{14}\,$G in accreting neutron star systems reduce electron scattering opacity for X-rays, resulting in an increase in the effective Eddington luminosity \citep{Herold1979,Eksie2018}. Strongly magnetized neutron stars ($B\geq 10^{14}\,G$) are required to produce the luminosities $\gtrsim 10^{40}\, {\rm erg\,s^{-1}}$ in this scenario \citep{Mushtukov2015}. 

\cite{Kluzniak-lasota2015} pointed out that ULX–2 M87 is not only distinguished from normal X-ray pulsars by its high luminosity but also the high spin-up rate of the neutron star that is two orders of magnitude higher than in normal X-ray pulsars. They showed that the spin-up rate can not be attributed to a magnetar.

In the KLK model (\citealt{king2017}, further studied in \citealt{king2017, king2020}), the high spin-up rate is explained by moderate strength magnetic fields (surface magnetic fields of $10^{10-11}\,$G) while the high luminosity is explained by beaming.

The scenario of accreting magnetars as ULXs was rejected by \cite{Lasota_2023}. They noted that the required magnetic field strength to increase the radiation luminosity up to the observed isotropic luminosity of ULXs is inconsistent with the spin-up rates seen in pulsating ULXs. 

In conclusion, beaming and a moderate magnetic field are required to explain the observed properties of ULXs powered by accreting neutron stars.

We investigate the impact of the moderately strong dipolar magnetic field in the range of $10^{10}\,$G to $10^{11}\,$G on the luminosity and beaming emission. In Section~\ref{Num}, we outline the numerical method and the used simulation setup. Simulation outcomes and discussion are given in Section~\ref{Res}. We summarize our findings in the concluding Section~\ref{Conc}.

\section{Numerical methods and simulations setup}\label{Num}

We use the \texttt{Koral} code \citep{Sadowski+dynamo,sadowski13}, to solve the equations of GRRMHD on a static mesh with a fixed metric. Our setup is the same as described in \cite{abarca+21}, briefly summarized as follows. Conservation equations for matter and radiation energy-momentum are solved separately using standard explicit methods for gas and the $M_1$ closure scheme for radiation. Matter and radiation energy-momentum tensors are coupled with the radiation four-force \citep{mihalas84} through a local implicit method. The magnetic field is evolved by using the flux-interpolated constrained transport method \citep{toth+00}, ensuring the divergence of the magnetic field remains at zero. The strong magnetizations within the neutron-star magnetosphere are addressed through a pioneering flooring scheme outlined by \citet{parfrey+17}, later expanded for radiation considerations by \cite{abarca+21}. 

%%%%%%%%%%%%%%%%%%%%%%%%%%%%%%%%%%%%%%%%%%%%%%
\begin{figure}
    \centering
    \includegraphics[width = \linewidth]{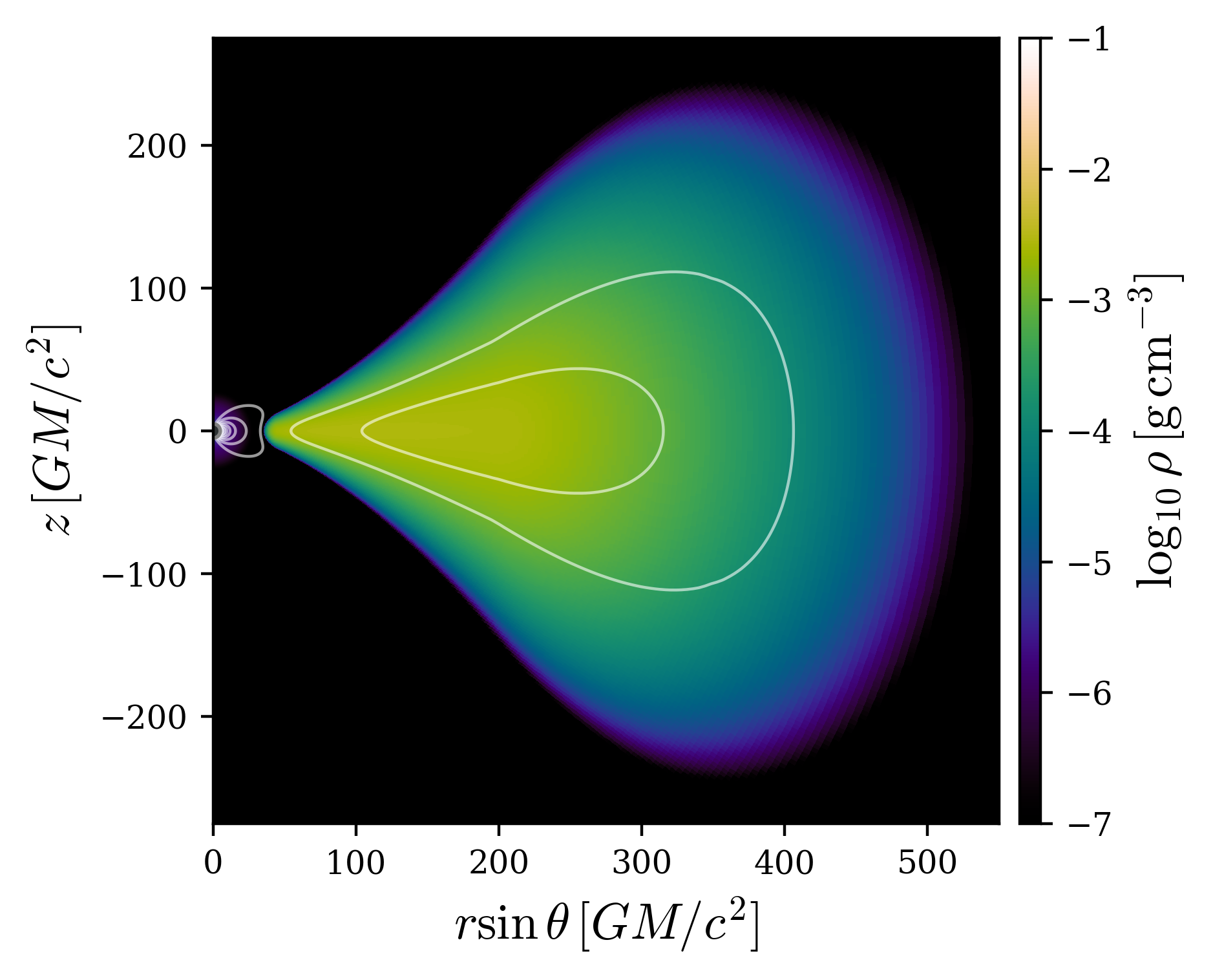}
    \caption{The initial rest mass density $\rho$ in the fluid frame, shown with logarithmic color grading. Solid lines indicate the isocontour lines of the vector potential $A_{\phi}$, which in our setup are parallel to the magnetic field lines. The surface magnetic field strength of dipole is $10^{11}\,$G.}
    \label{init}
\end{figure}
%%%%%%%%%%%%%%%%%%%%%%%%%%%%%%%%%%%%%%%%%%%%%%%%%%%%%%%%%%%%%%%%%%%%%%
%%%%%%%%%%%%%%%%%%%%%%%%%%%%%%%%%%%%%%%%%%%%%%%%%%%%%%%%%%%%%%%%%
\begin{figure*}
    \centering
    \includegraphics[width=1\linewidth]{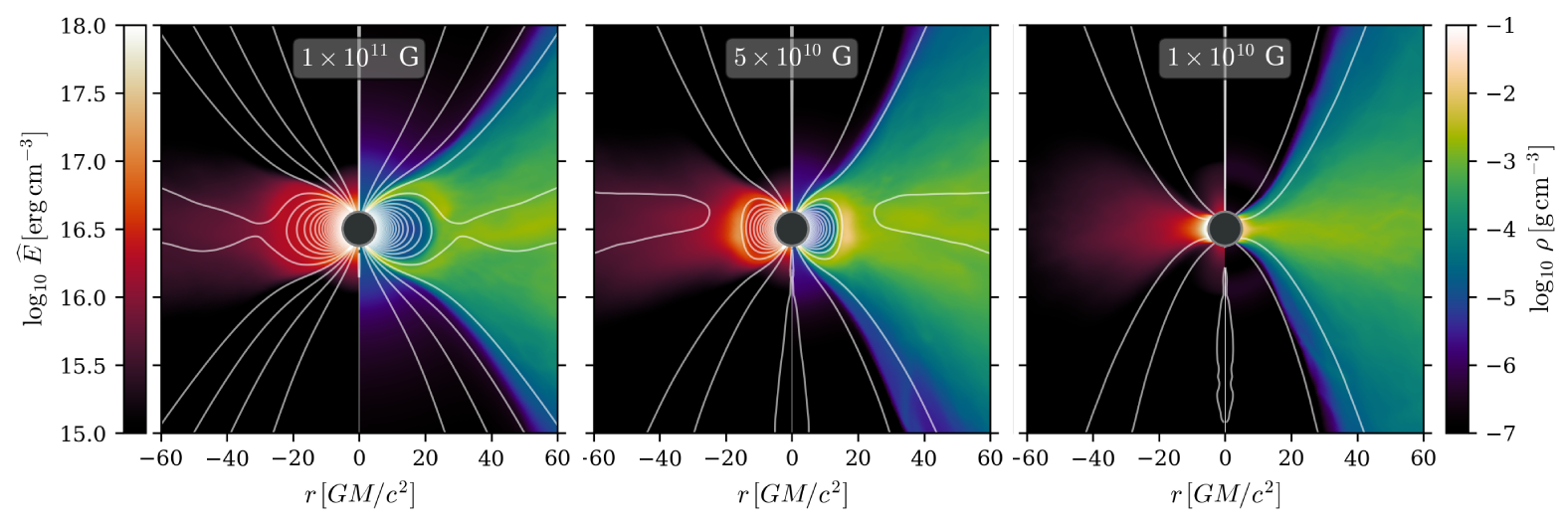}
    \caption{Our simulation results in the cases with stellar dipoles of $10^{11}\,$G (\textit{left panel}), \fk{$5\times10^{10}\,$G (\textit{middle panel})} and $10^{10}\,$G (\textit{right panel}). The left half of the panels shows the radiation energy density $\hat{E}$ and the right half the rest-mass density $\rho$. Magnetic field lines are plotted as isocontours of the vector potential, $A_{\phi}$. Note that the rarefied cones above and below the star are much wider for the $10^{11}\,$G case. The plots are produced using the time-averaged data.}
    \label{avgrhorad}
\end{figure*}
%%%%%%%%%%%%%%%%%%%%%%%%%%%%%%%%%%%%%%%%%%%%%%%%%%%%%%%%%%%%%%%%%%
%%%%%%%%%%%%%%%%%%%%%%%%%%%%%%%%%%%%%%%%%%%%%%%%%%%%%%%%%%%%%%%%%%%%%
\begin{figure*}
    \centering
    \includegraphics[width=1\linewidth]{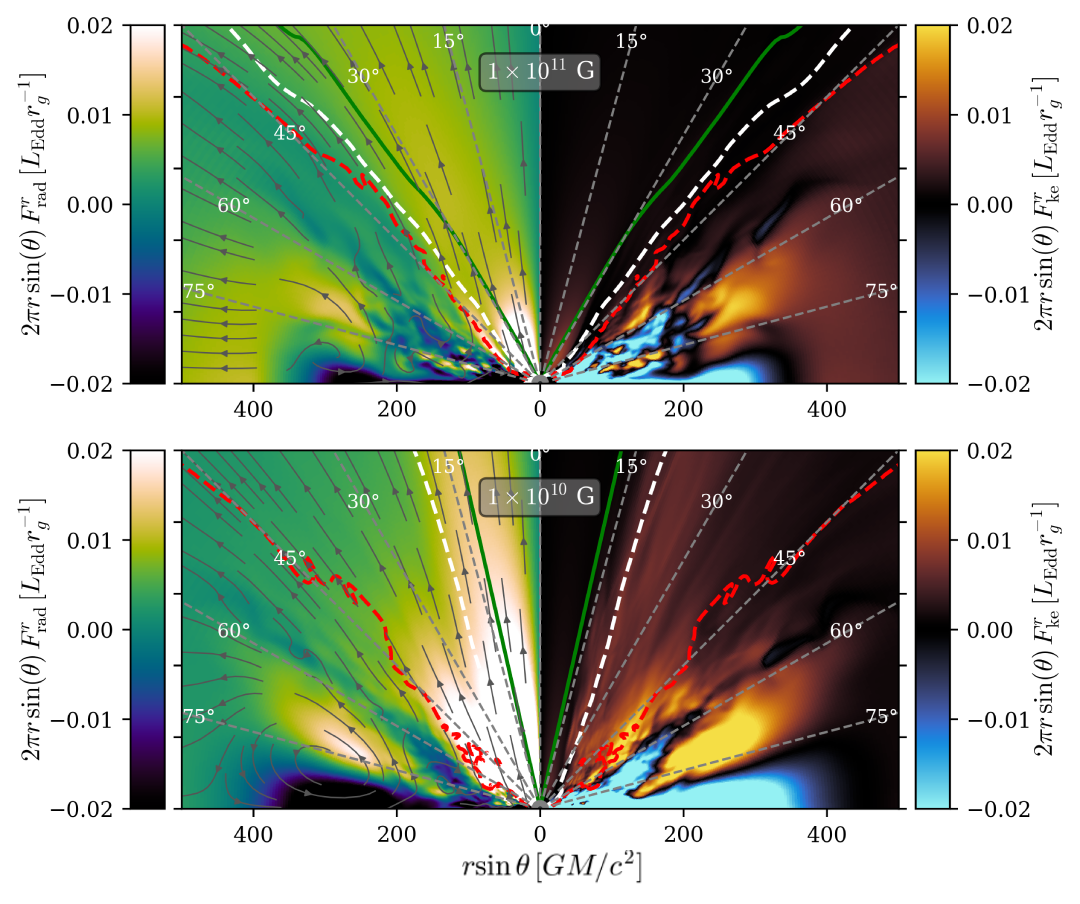}
    \caption{The radiation and kinetic fluxes in our simulations with the neutron star dipole $10^{11}\,$G (\textit{top panel}) and $10^{10}\,$G (\textit{bottom panel}) are shown in the left and right halves of each panel, respectively. The negative values indicate the direction towards the neutron star (inflow), while the positive values indicate outflow. The red dashed contour represents the zero Bernoulli surface. The white dashed and solid green contours depict photospheres $\tau_\theta = 1$ and  $\tau_r = 1$, respectively. The dashed grey lines show the viewing angles as labeled. The grey streamlines in the left half panels follow the radiation flux and indicate its direction. The plots are produced using the time-averaged data.}
    \label{outflow}
\end{figure*}
%%%%%%%%%%%%%%%%%%%%%%%%%%%%%%%%%%%%%%%%%%%%%%%%%%%%%%%%%%%%%%%%%%%%%

We use a 2.5D grid based on the Schwarzschild metric with the signature $(-, +, +, +)$. The resolution of the simulations is $512 \times 510 \times 1$ cells in $r$, $\theta$ and $\phi$ directions, respectively. The grid spacing is logarithmically increasing in a radial direction, spanning from $r_{\rm in}=5 \, r_{\rm g}$ to $r_{\rm out}=1000\, r_{\rm g}$. It is a well-established fact that in the axisymmetric cases, the magnetic field inside the accretion disk decays over time \cite[the anti-dynamo theorem described in][]{cowling+33}. To address this, we utilize a mean-field dynamo \citep{Sadowski+dynamo} which effectively restores the magnetic field, mimicking the behavior expected in a full 3D simulation. This approach enables us to conduct long-duration simulations. We run the simulations to $50\,000 \, t_{\rm g}$, where $t_{\rm g} = GM/c^3$ is the gravitational time. We use the units where $G = c = 1$ in the equations. 

We conducted simulations employing different strengths of the stellar dipole magnetic field, $1\times 10^{10}$, $5\times 10^{10}$, and $1\times 10^{11}\,$G. The neutron star mass is $1.4\,M_\odot$ and the radius is $5\, r_{\rm g}$. We neglect the stellar rotation because the simulations run for $50\,000$ $t_{\rm g}$, which is about $0.35 \,{\rm s}$ of physical time, much smaller than the period of observed pulsating ULXs, which is on the order of seconds. All the simulations are initialized with the same equilibrium torus, which produces a mass accretion rate beyond $200\, L_{\rm Edd}c^{-2}$. \fk{The torus is weakly magnetized with a single-loop magnetic field of initial strength $B = 10^{-12}\,$G.} 

\fk{We apply the energy-reflecting boundary condition 
\citep{abarca+21} on the neutron star surface, where the thermal and kinetic energy of the interacting gas with the neutron star surface is released as radiation energy. We also permit a portion of this radiation energy to be absorbed by the neutron star surface. In the simulations presented in this paper, we utilize an albedo of 0.75, representing a fraction of the radiation energy released.}

% with the ratio of the total pressure of gas and radiation to the magnetic pressure $\beta = (p_{\rm gas} + p_{\rm rad})/p_{\rm mag} = 10$.
The initial setup for the simulation with the dipolar magnetic field of $10^{11}\,$G is shown in Fig.~\ref{init}. The simulation is initialized with a background atmosphere with density and internal energy several orders lower than in the initial torus. The inner radius of the torus is $34\,r_{\rm g}$ and the outer radius is $500\,r_{\rm g}$. We also note that in our model the disk is (nearly) Keplerian.

%%%%%%%%%%%%%%%%%%%%%%%%%%%%%%%%%%%%%%%%%%%%%%%%%%%%%%%%%%%%%%%%%%%%%
\section{Results and discussion}\label{Res}
We present the results in our simulations with the time-averaged data over the periods from $t=15\,000 \,t_{\rm g}$ to $50\,000 \,t_{\rm g}$.  

In Fig.~\ref{avgrhorad} are shown the results for the radiation energy density and the rest-mass density, for \fk{three} simulations with the neutron star dipole magnetic fields \fk{$10^{11}$, $5\times10^{10}$, and $10^{10}\,$G from left to right}. The disk is truncated at the magnetospheric (\alfven) radius where the magnetic pressure dominates the ram pressure. The magnetospheric radius increases with the strength of the magnetic field. In the case of the weak magnetic field ($10^{10}\,$G) it is at $5.3\,r_{\rm g}$, very close to the surface of the neutron star, \fk{while in the simulation with the stronger magnetic fields $B = 5\times 10^{10}\,$G and $10^{11}\,$G it is at about $13 \,r_{\rm g}$ and $18 \,r_{\rm g}$, respectively.} 

During the simulation, the stellar dipole and the torus field lines, which are oriented in opposite directions, converge and undergo reconnection \citep{parfrey+17,abarca+21}. As evident in the mid-plane of the disk of the left frame in Fig.~\ref{avgrhorad}, in the presence of the strong magnetic field ($10^{11}\,$G) the dipolar field lines close to the neutron star remain unchanged and the stream of dense matter follows the disk mid-plane field lines and the accretion column and falls on the surface of the neutron star. The right frame of Fig.~\ref{avgrhorad} shows that in the simulation with the weaker dipole ($10^{10}\,$G), almost all dipolar field lines break, and gas propagates to higher latitudes. Thus, the radiation energy becomes more tightly collimated towards the polar axis. We discuss this fact in more detail later in this section.

Fig.~\ref{avgrhorad} shows that where the accreting material reaches the surface of the neutron star in the simulation with the weak magnetic field (right frame), radiation freely propagates toward the observer along angles close to the polar axis. In the simulation with the strong magnetic field (left frame), there is a thin layer of low-density material above the neutron star surface, between the accretion column and the neutron star axis. It is caused by the trapping of the background atmosphere in the strong magnetic field and this layer may slightly affect the radiation luminosity. 

As the gas moves towards the neutron star, it releases energy in the form of thermal, kinetic and radiation energies. In the accretion process, gravitational potential energy is converted into kinetic energy. Collisions within the disk convert kinetic energy into thermal energy, which heats the disk and causes radiation emission. 
It is crucial to accurately compute the kinetic energy that accelerates the outflow and powers the radiation luminosity. The radiation flux is 
\begin{equation}
    F^r_{\rm rad}= - R^r\phantom{}_t,
\end{equation}
 where $R^r \phantom{}_t$ is the radial component of radiation energy-momentum tensor, and the kinetic flux is 
\begin{equation}\label{ke_flux}
 F^r_{\rm ke} = - \rho u^r(u_t + \sqrt{-g_{tt}}) ,
\end{equation}
where $u_t$ is the time component of 4-velocity, and $g_{tt} = -(1-2r_{\rm g}/r)$ in the Schwarzschild metric \citep[for details see][]{sadowski2016}. With this signature, $u_t < 0$. 

In Fig.~\ref{outflow} are shown the radiation flux (left-half panels) and kinetic flux (right-half panels) for two magnetic fields $10^{11}\,$G (in the top panel) and $10^{10}\,$G (in the bottom panel). The fluxes are multiplied by $ 2\pi r \sin \theta$ and shown in the units of [$L_{\rm Edd} r^{-1}_{\rm g}$]. In each frame, the radius is extended to $500 r_{\rm g}$, where the outer edge of the torus is located. 

Three contours in Fig.~\ref{outflow} separate different regions. The red dashed contour shows the zero Bernoulli surface where the relativistic Bernoulli parameter has the value $Be = 0$, with 
\begin{equation}
    Be = -\, \frac{T^t \phantom{}_t + R^t\phantom{}_t + \rho u^t}{\rho u^t},
\end{equation}
with $T^t\phantom{}_t$  and $R^t\phantom{}_t$ being the MHD and radiation energy densities, respectively. The zero Bernoulli surface splits the simulation domain into two parts. Above this surface, the gas with $Be > 0$ is energetic enough to reach infinity on its own, without an external source of energy. Below the surface, with $Be < 0$, the gas is gravitationally bound. The white dashed and solid green lines represent surfaces of the photosphere computed in $\theta$ and $r$ directions, respectively. The scattering optical depth along these contours are $\tau_{\theta} = 1$ and $\tau_{\rm r} = 1$, which are measured as
%%%%%%%%%%%
\begin{equation}
        \tau_\mathrm{\theta}(\theta) = \int_{0}^{\theta} \rho \kappa_{\mathrm{es}} \sqrt{g_{\rm \theta \theta}}d{\theta},
\end{equation}
%%%%%
\begin{equation}
        \tau_\mathrm{r} (r)= \int_{r}^{r_{\rm out}} \rho \kappa_{\mathrm{es}} \sqrt{g_{\rm rr}} d{r},
\end{equation}
%%%%
where $\kappa_{\rm es} = 0.34 \,{\rm cm}^2\,{\rm g}^{-1}$ is the electron scattering opacity for solar composition and $r_{\rm out}$ the outer boundary of the simulation.

The top-right panel in Fig.~\ref{outflow} shows that in the strong magnetic field simulation ($10^{11}\,$G) the outgoing kinetic flux above the zero Bernoulli surface is negligible. In the weak magnetic field simulation ($10^{10}\,$G), shown in the bottom-right panel, above zero Bernoulli surface there is a significant amount of outgoing kinetic flux. The outflows impact the photosphere structure in such a way that the photosphere surfaces $\tau_\theta =1$ and $\tau_{\rm r}=1$ (shown with the white dashed and solid green lines, respectively) are close to the polar axis of the neutron star in the weak magnetic field simulation. The photospheric surface $\tau_{\rm r} = 1$ is located at the angle of $35^\circ$ in the simulation with magnetic field $10^{11}\,$G, while it is along the viewing angle of about $15^\circ$ in the simulation with magnetic field $10^{10}\,$G.  

The radiation flux is shown in the left half-panels of Fig.~\ref{outflow}. The radiation flux is more beamed in the weak magnetic field simulation compared to the strong magnetic field simulation, attributable to the optically thin region ($\tau_{\rm r}<1$) being narrower. 

The top-left panel of Fig.~\ref{outflow} shows that in the strong magnetic field simulation, there is outgoing radiation flux within the zero Bernoulli surface at $r\gtrsim 300 \, r_{\rm g}$. However, the simulation at the radii above $50\,r_{\rm g}$ contains numerical artefacts because the disk has not converged within the simulation time. 

We find that in the simulation with the strong magnetic field, there is a larger amount of ingoing radiation flux along the equatorial plane, at the radii less than 100~$r_{\rm g}$, compared to the simulation with the weak magnetic field. 

\begin{figure}
    \centering
    \includegraphics[width = \linewidth]{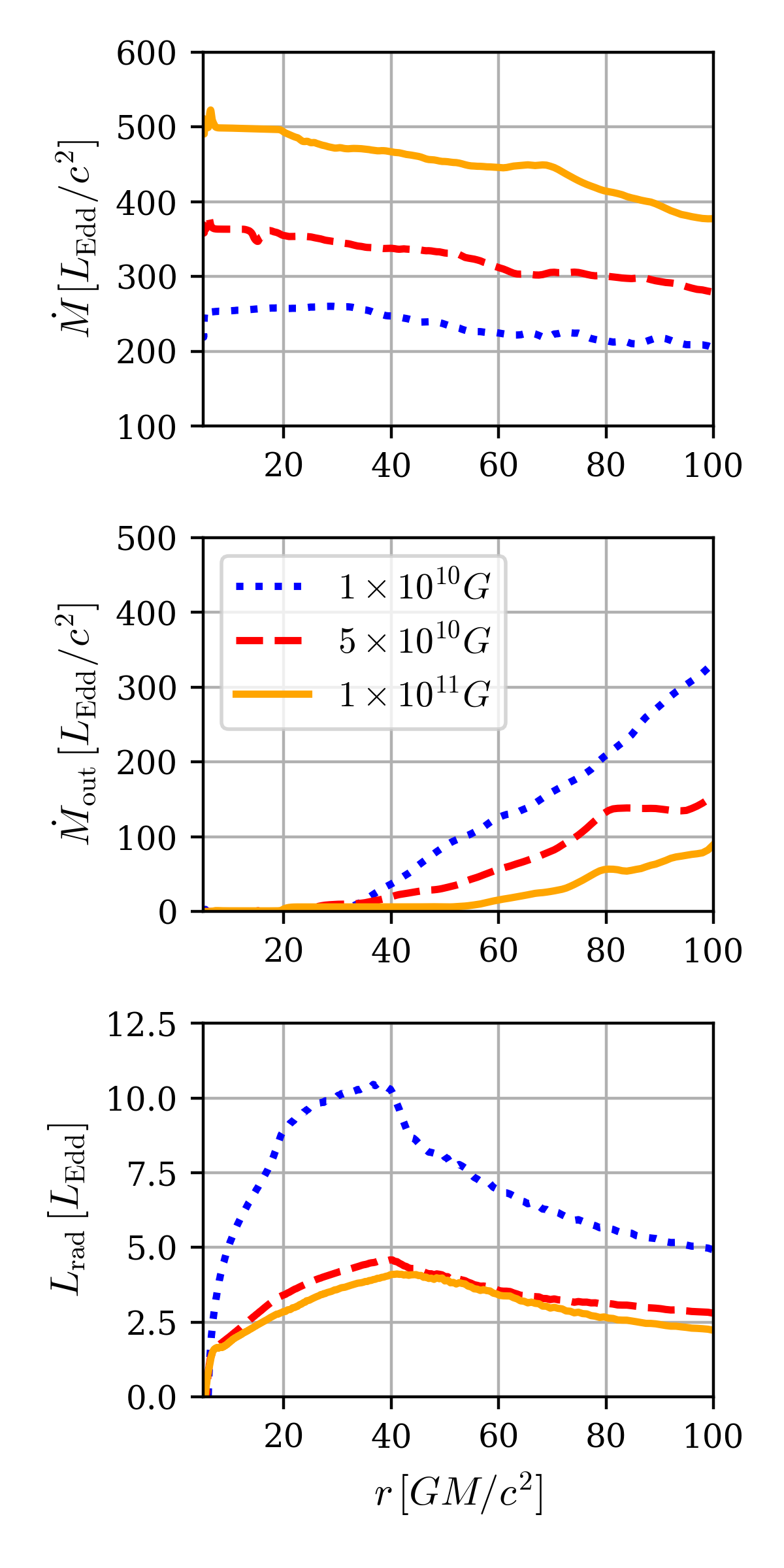}
        \caption{The accretion rate $\dot{M}$ (\textit{top panel}) and outflow rate $\dot{M}_{\rm out}$ (\textit{middle panel}) are shown in the unit of  $L_{\rm Edd}/c^2$. \textit{Bottom panel:} radiation luminosity in the optically thin region $\tau_{\theta}<1$ in the unit of $L_{\rm Edd}$. Results of three different simulations are shown, with the magnetic fields as in the legend of the middle panel.}
    \label{mdots}
\end{figure}

The accretion rate and outflow for three simulated magnetic field models are shown in Fig.~\ref{mdots}. The accretion rate $\dot{M}(r)$ is measured by integrating the momentum density $\rho u^r$ over a spherical surface of a given radius: 
%%%%
\begin{equation}
    \dot{M}(r) = - 2\pi\int_{0}^\pi \rho u^r r^2 \sin\theta \, d{\theta},
\end{equation}\label{mdotin}
%%%%
and the outflow rate $\dot{M}_{\rm out}$ is computed from the momentum density of the gas flowing away from the accretor ($u^r>0$):
\begin{equation}
    \dot{M}_{\rm out}(r)  = 2 \pi\int_{0}^\pi \rho u^r r^2 \sin\theta \, d{\theta}\bigg|_{u^r>0}.
\end{equation}\label{mdotout}
%%%%
The radius at which the accretion rate is constant is called the steady-state radius. In our simulations, the accretion rate reaches the steady-state at $\sim 30 \, r_{\rm g}$ in the simulations with the magnetic fields $10^{10}\,$G and $5\times 10^{10}\,$G and at $20 \, r_{\rm g}$ in the simulation with the magnetic field $10^{11}\,$G. In our simulations, the accretion rate decreases with decreasing magnetic field. The middle panel of Fig.~\ref{mdots} shows that the outflow rate $\dot{M}_{\rm out}$ increases significantly with decreasing the strength of the dipole.
% \textbf{The significant outflows in the simulation with the weaker dipole ($10^{10}\,$G), can be attributed to the accreting material slowing down as it approaches the neutron star.}
%%%%%%%%%%%%%%%%%%%%%%%%%%%%%%%%%%%%%%%%%%%%%%%%%%%%%%%%%%%%%%%%%%%%%%%%
%%%%%%%%%%%%%%%%%%%%%%%%%%%%%%%%%%%%%%%%%%%%%%%%%%%%%%%%%%%%%%%%%%%%%%%%
Next, we compute the radiation luminosity in the optically thin region. Since radiation escapes from the accretion column along the $\theta$ direction, we compute the luminosity by integrating the radiation flux $F^r_{\rm rad}$ over the spherical shell above the photosphere with $\tau_{\theta} = 1$: 
%%%
\begin{equation}\label{radlum}
    L_{\rm rad} (r) =  2\pi \int_{\tau_\theta <1} F^r_{\rm rad} r^2 \sin \theta\, d{\theta}.
\end{equation}
%%%
The radiation luminosity within $\tau_\theta<1$ is displayed in the bottom panel of Fig.~\ref{mdots}. Due to the radiation shock, there is a steep rise with radius in radiation luminosity at the radii close to the star. Radiation luminosity increases up to radius $40\,r_{\rm g}$ as the radiation from the accretion column and disk are included. At distances greater than $40\, r_{\rm g}$, radiation decreases with increasing radius. This decrease in radiation luminosity is attributed to the momentum transfer of photons to the outflowing gas when they pass through the outflow region. In the simulation with the weakest magnetic field ($10^{10}\,$G) which shows larger outflows, the radial drop in luminosity is more noticeable compared to the stronger magnetic field simulations. 

We calculate the luminosity that reaches the observer by computing the luminosity in the optically thin region within the optical depth $\tau_{\rm r} < 1$, 
%%%
\begin{equation}\label{radlum}
    L (r) =  2\pi \int_{\tau_r <1} F^r_{\rm rad} r^2 \sin \theta\, d{\theta}.
\end{equation}
%%%
Since the accretion rate in each of the simulations is different, to enable comparison, we compute the radiation efficiency defined as the ratio of the radiation luminosity to the accretion rate $\dot{M}c^2$. The accretion rate is calculated at a radius of $20 \, r_{\rm g}$. As mentioned earlier, at this radius, we find that the disk is in a steady-state in all simulated models we presented here. 

\begin{figure*}
    \centering
    \includegraphics[width = 0.9\linewidth]{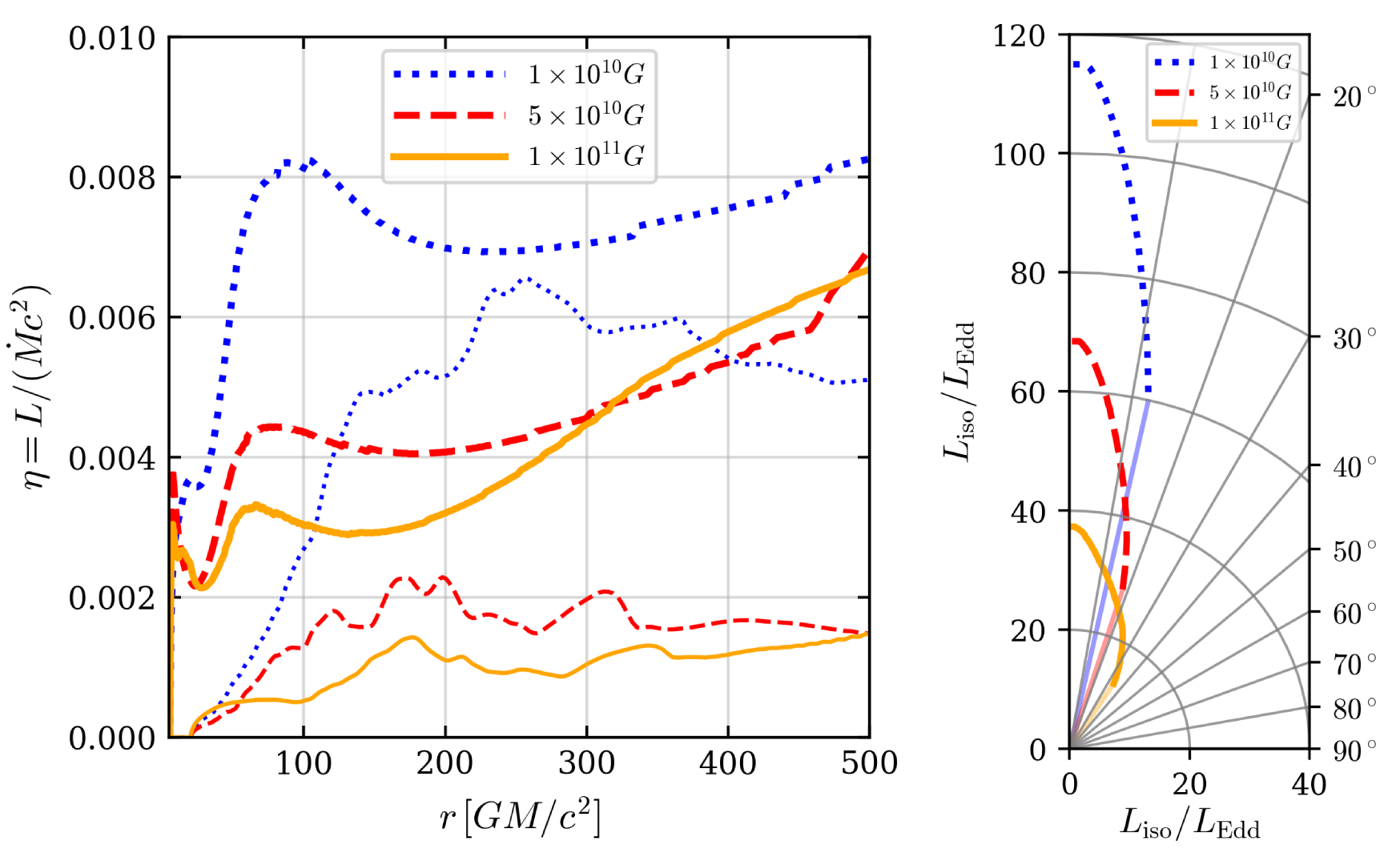}
    \caption{{\sl Left panel:} The radiative efficiency in the optically thin region $\tau_{\rm r} <1$ (thick lines) and kinematic efficiency (thin lines) both relative to the mass accretion rate onto the neutron star, $L/(\dot{M}c^2)$, as a function of the radius. The kinetic luminosity is computed from outgoing kinetic flux above the torus $\theta \leq60^\circ$ shown in Fig.~\ref{outflow}.
    {\sl Right panel:} The isotropic luminosity as a function of the viewing angle, $L_{\rm iso}(\theta) = 4\pi d^2 F^r_{\rm rad}(\theta)$, computed at $d=500 r_{\rm g}$ in the optically thin cone, $\tau_{r} < 1$. The radiation luminosity is taken to drop to zero at $\tau_{r} = 1$ shown with faded colors, i.e. at viewing angles corresponding to the opening angle of the optically thin cone. In this polar diagram of the beaming pattern, the straight grey solid lines correspond to particular viewing angles. The magnetic field in each simulation is shown with a distinctive color and line-styles.}
    \label{apparent}
\end{figure*}

The left panel of Fig.~\ref{apparent} shows the efficiency of radiation $L_{\rm rad}/(\dot{M}c^2)$ represented by thick colored lines, with different colors and line-styles corresponding to different magnetic field simulations. As the magnetic field increases, the efficiency of radiation luminosity decreases. A peak in the curve close to the star is caused by the radiation shock which is more significant in the stronger magnetic field simulations. We note that although we have implemented an energy-reflecting boundary condition with a maximal albedo of 0.75 \citep{abarca+21, dabarca_thesis}, the photon-trapping phenomenon \citep{sadowskiNarayan2016, Ohsuga2002} is so strong that the advected radiation flux dominates energy transport close to the surface of the neutron star. The actual albedo turns out to be $\simeq 0.1$. Consequently, the radiation efficiency is much lower than expected. The fluctuation in the curves at the small radii is most likely caused by the geometry of the photosphere close to the neutron star.  After a rapid increase with radius out to $\sim 100\, r_{\rm g}$, the radiation efficiency curve flattens at different radii depending on the strength of the dipole. At the radii where the curves are flat, the radiation efficiency of the simulation with the weakest magnetic field ($10^{10}\,$G) is about 0.007, while in the simulation with the strongest magnetic field ($10^{11}\,$G), is about 0.003. In the simulation with the weakest magnetic field ($10^{10}\,$G), there is a slight increase in the luminosity at $\gtrsim 400\, r_{\rm g}$, while in the strong magnetic field simulations ($\geq 5\times10^{10}\,$G) the luminosity significantly increases at the radii $\gtrsim 200\, r_{\rm g}$. This increase is most likely due to the curvature of the photosphere at large radii allowing the radiation to escape to the optically thin region (see Fig. \ref{outflow}). The curvature of the photosphere can be attributed to less outflows of matter at larger radii.

% the gas becoming thinner and allowing more radiation to escape into the optically thin region. It might be also caused by the geometry of the photosphere at larger radii in the simulation with the strong magnetic field, as shown in Fig. \ref{outflow}. 

The power in the outflow is computed by integrating the outgoing kinetic flux (given in Eq. \ref{ke_flux}) over a spherical surface at each radius,
\begin{equation}
    L_{\rm ke} (r) = 2\pi  \int_{0}^{\theta=60^\circ} F^r_{\rm ke} r^2 \sin(\theta)\, d{\theta} \bigg|_{u^r>0},
\end{equation}
where the angle $\theta= 60^\circ$ is chosen to exclude the energy evolution within the torus.
%%%%%%%%%%%%
\begin{figure*}
    \centering
    \includegraphics[width = 0.9\linewidth]{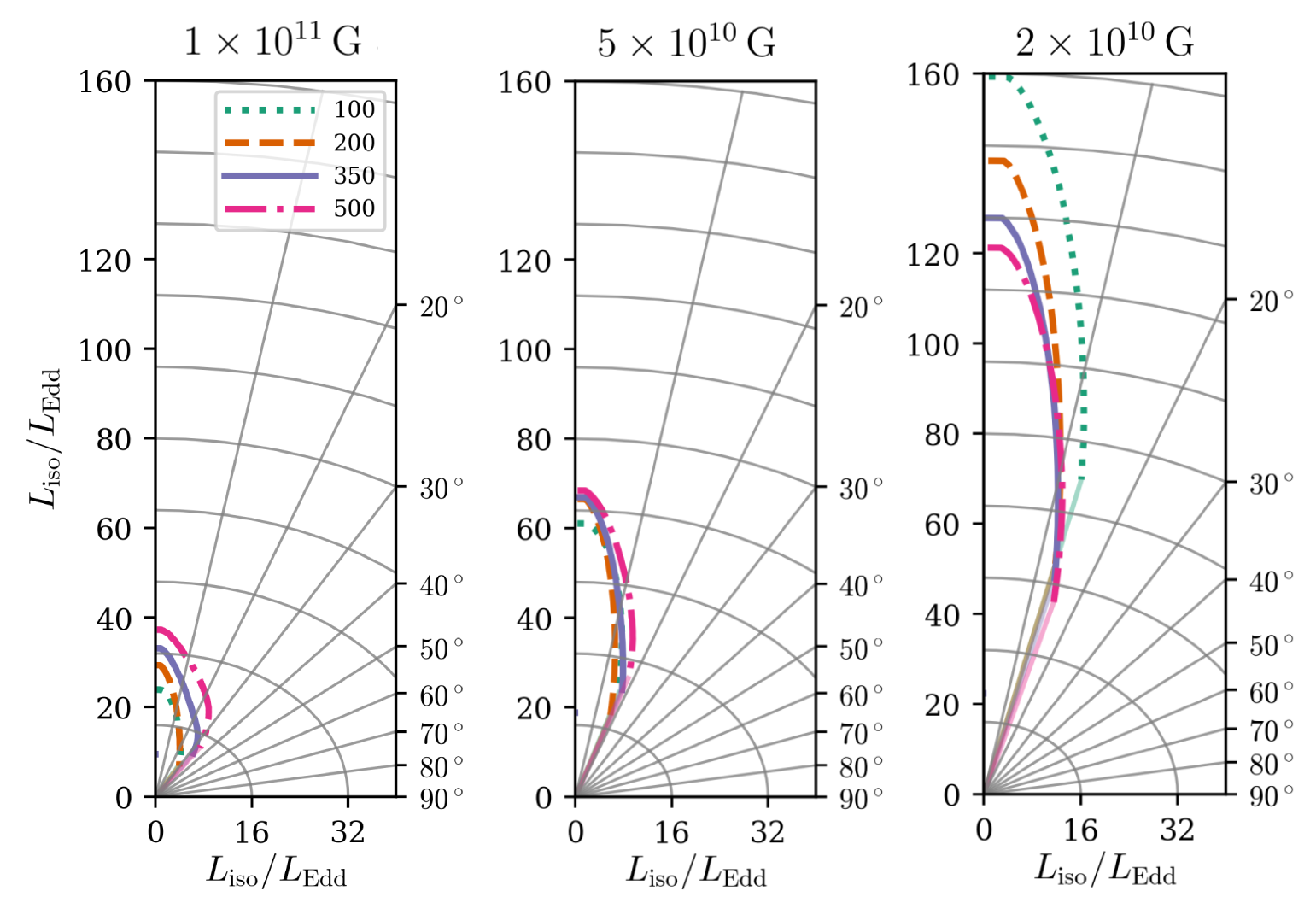} 
    \caption{\fk{The apparent luminosity as a function of the viewing angle, $L_{\rm iso}(\theta) = 4\pi d^2 F^r_{\rm rad}(\theta)$, computed at radii 100, 200, 350, and 500$\,r_{\rm g}$ for three magnetic dipole strengths $10^{11}\,$G in the {\sl left panel}, $5\times 10^{10}\,$G in the {\sl middle panel}, and $2\times 10^{10}\,$G in the {\sl right panel} \citep[the plot in the right panel is produced using the simulation data in][]{abarca+21}.}}
    \label{appar_dist}
\end{figure*}
%%%%%%%%%%%%
The efficiency of the kinetic luminosity $L_{\rm ke}/(\dot{M}c^2)$ of the outflowing gas in each magnetic field simulation is represented by the three thin lines in the left panel of Fig.~\ref{apparent}. The kinetic efficiency is of the same order of magnitude as the radiation efficiency in the simulation of the weakest magnetic field ($10^{10}\,$G) and about half of the radiation efficiency in the strongest magnetic field simulations ($10^{11}\,$G). At radii less than $40\, r_{\rm g}$, where there are no outflows (as shown in the middle panel of Fig.~\ref{mdots}) the kinetic luminosity diminishes to zero. At distances beyond $40\, r_{\rm g}$ with a significant amount of the outflow the kinetic luminosity increases steeply. In the simulations with dipole strengths $\geq 5\times 10^{10}\,$ G the kinetic efficiency is almost constant ($\leq0.002$) beyond the radius of about $100\, r_{\rm g}$. In the simulation with the weak magnetic field ($10^{10}\,$G) there is a steep rise in the kinetic efficiency to the radius of $150\, r_{\rm g}$, continuing to $250\, r_{\rm g}$. The maximum kinetic efficiency with this magnetic field simulation is about 0.007. 

From the observational perspective, isotropic (apparent) luminosity $L_{\rm iso}$ is an important parameter. But if the source is not isotropic, the observer overestimates $L_{\rm iso}$ by a factor of $1/b$.    

It was shown through numerical simulations
in \cite{abarca+18} that non-magnetized neutron stars cannot have an apparent luminosity that exceeds the Eddington limit, even with a super-Eddington accretion rate. Subsequently, \cite{abarca+21} showed that in the simulation of a super-Eddington accreting neutron star with a dipolar magnetic field $2 \times 10^{10}\,$G, the apparent luminosity is beyond $100\, L_{\rm Edd}$.

We compute the apparent luminosity $L_{\rm iso}= 4 \pi d^2 F^r_{\rm rad}$ where $d$ is the distance between an observer and the object. In the right panel in Fig.~\ref{apparent} we show the apparent luminosity at $ d =  500\,r_{\rm g}$. To include all the radiation luminosity that may reach the observer depending on their viewing angle (neglecting cosmological effects) we compute $F^r_{\rm rad}$ in the optically thin region $\tau_{\rm r} <1$. The apparent luminosity $L_{\rm iso}$ is shown in the units of $L_{\rm Edd}$ in a polar diagram. The viewing angles are indicated with straight grey solid lines. At large angles, where the optically thick disk is located, no radiation flux can be detected by the observer, so the apparent luminosity is zero. In our simulations, this angle changes with respect to the strength of the dipole which is shown by the faded-color lines for each magnetic field simulation. For instance, if we look at the object with a magnetic field $10^{10}\,$G from a viewing angle of about $15^\circ$ or more, the apparent luminosity of the accreting neutron star is zero.
The apparent luminosity reaches the maximum value along the polar axis. It is seen that the apparent luminosity decreases with increasing dipole strength, although the accretion rate changes conversely. 
The apparent luminosity reaches about $120\,L_{\rm Edd}$ for the simulation with dipole strength $10^{10}\,$G, and is only about $40\,L_{\rm Edd}$ in the simulation with dipole strength $10^{11}\,$G. Our simulations show that ULXs may well be powered by accreting neutron stars with a dipole of $10^{10}\,$G. However, the luminosity we find is on the lower side luminosities of detected ULXs.

\section{Summary and conclusions}\label{Conc}

We study beamed X-ray emission from neutron stars in the context of ULXs. We perform general relativistic radiative magnetohydrodynamics simulations of supercritical accretion onto neutron stars with dipolar magnetic fields in the range of $10^{10}$ to $10^{11}\,$G (maximum surface values, corresponding to dipole strengths $10^{28}$ to $10^{29}\,{\rm G\cdot cm^3}$ for our neutron star radius of about 10~km). The mass accretion rate onto the star is set to be above $200\, L_\mathrm{Edd}c^{-2}$; the rate of mass flow through the accretion disk is even higher, reflecting strong outflows from the disk. 

Our simulations show that the magnetic field of $10^{10}\,$G leads to significant mass outflows of about $300\, L_{\rm Edd}c^{-2}$ in kinetic energy, locating the photosphere close to the polar axis of the neutron star, roughly on the surface of a cone of opening angle $\approx 15^{\circ}$. Thus, the radiation escaping toward the observer through the optically thin cone is highly beamed and reaches an apparent luminosity of about 120 Eddington units along the polar axis although the true luminosity is about 10 Eddington units. For a magnetic field increased by an order of magnitude (B = $10^{11}\,$G) the outflow is about $ 100\, L_{\rm Edd}c^{-2}$, and the optically thin region is widened to an angle of about $35^\circ$. Here, the luminosity is less collimated compared to the lower magnetic field. The maximum value of apparent luminosity is about 40 Eddington luminosity along the polar axis. 

The beaming factor $b$ is about 0.02 in the weak magnetic field simulation and about 0.08 in the strong magnetic field simulation. 
The beaming factor in our model is comparable to the one estimated in the KLK model \citep{king2017, king2019, king2020}. However, our simulations cannot be directly compared with the KLK model, as the spherization and magnetospheric radii are close to each other in the KLK model, whereas, in our simulations, the spherization radius is significantly larger. 

\fk{We calculate the apparent luminosity at various distances from the source. It is shown that in the simulation with the magnetic field $10^{11}\,$G the apparent luminosity increases with increasing distance from the source, which is opposite to the simulation with a magnetic field strength of $2 \times 10^{10}\,$G \citep[from][]{abarca+21}.} 

We conclude that ultraluminous X-ray sources are most likely powered by neutron stars with a dipolar magnetic field in order of $10^{10}\,$G than with a stronger magnetic field.

The caveat of our simulations is the inner boundary condition. Although we use the energy-reflecting neutron star surface with an albedo of 0.75 the actual value turns out to be about 0.1 which leads to the radiation efficiency being much less than the expected value. The small value of albedo might be a numerical artefact and/or caused by the extreme condition of the simulation such as a high accretion rate. More simulations with various albedo and accretion rates are required to examine the energy-reflecting boundary condition.  

\fk{We also note that although the simulation time is long enough that the disk converges out to radius 20-40$\,r_{\rm g}$, the simulation time is about one-third of the period of neutron star ULXs so that the spin of neutron star is ignored. Considering the spin of the neutron star in 2-dimensional simulation does not provide more insight into the understanding of neutron star ULXs while in 3-dimensional simulation it gives the possibility to study pulsation of the neutron star ULXs. It is also notable that in 3-dimensional simulations the Rayleigh-Taylor instability is maintained in the closed field lines zone, as discussed in \cite{parfrey2023,Romanova2008}. In our 2.5D simulations, the instability in the closed zone only appears briefly. As we use the time-averaged
data over a long period, it doesn’t have a significant effect on the results we report. However, only 3D simulations will be able to determine the importance of the Rayleigh-Taylor instability in radiative simulations.}

In future work, \fk{we will study the pulsating ULXs through 3-dimensional simulations}. We will also conduct numerical simulations across a wide range of magnetic fields to determine the limits in interpreting ultraluminous X-ray sources powered by accreting neutron stars. We also note that more accurate computation of the luminosity and post-processing radiation transfer would be needed.

\section*{Acknowledgments}
F.K. thanks David Abarca and Dominik Gronkiewicz for their useful discussions and technical help, and Jean-Pierre Lasota and Agata Różańska for their valuable advice and suggestions. 
Research in CAMK was in part supported by the Polish National Center for Science grant 2019/33/B/ST9/01564. F.K. acknowledges the Polish National Center for Science grant no. 2023/49/N/ST9/01398. M.Č. acknowledges the Czech Science Foundation (GAČR) grant No. 21-06825X and the support by the International Space Science Institute (ISSI) in Bern, which hosted the International Team project No. 495 (Feeding the spinning top) with its inspiring discussions. 
We gratefully acknowledge Polish high-performance computing infrastructure PLGrid (HPC Center: ACK Cyfronet AGH) for providing computer facilities and support within computational grant no. PLG/2023/016648.

\bibliography{NSbeaming}{}
\bibliographystyle{aasjournal}

\end{document}